\def\pr#1 {Phys. Rev. {\bf D#1\tie \rm }}
\def\pe#1 {Phys. Rev. {\bf #1\tie\rm }}
\def\pre#1 {Phys. Rep. {\bf #1\tie\rm }}
\def\pl#1 {Phys. Lett. {\bf #1B\tie \rm }}
\def\prl#1 {Phys. Rev. Lett. {\bf #1\tie \rm }}
\def\np#1 {Nucl. Phys. {\bf B#1\tie \rm }}
\def\ap#1 {Ann. Phys. (NY) {\bf #1\tie \rm }}
\def\cmp#1 {Commun. Math. Phys. {\bf #1\tie \rm }}
\def\imp#1 {Int. Jour. Mod. Phys. {\bf A#1\tie \rm }}
\def\mpl#1 {Mod. Phys. Lett. {\bf A#1\tie\rm }}
\def\jhep#1 {JHEP {\bf #1\tie\rm }}
\def\zp#1 {Z. Phys. {\bf C#1\tie\rm }}

\def\tie{\noexpand~}

\def\be{\begin{equation}}
\def\ee{\end{equation}}
\def\bea{\begin{eqnarray}}
\def\eea{\end{eqnarray}}

\catcode`@=11
\def\marginnote#1{}
\newcount\hour
\newcount\minute
\newtoks\amorpm
\hour=\time\divide\hour by60
\minute=\time{\multiply\hour by60 \global\advance\minute by-\hour}
\edef\standardtime{{\ifnum\hour<12 \global\amorpm={am}%
        \else\global\amorpm={pm}\advance\hour by-12 \fi
        \ifnum\hour=0 \hour=12 \fi
        \number\hour:\ifnum\minute<10 0\fi\number\minute\the\amorpm}}
\edef\militarytime{\number\hour:\ifnum\minute<10 0\fi\number\minute}
\def\draftlabel#1{{\@bsphack\if@filesw {\let\thepage\relax
   \xdef\@gtempa{\write\@auxout{\string
      \newlabel{#1}{{\@currentlabel}{\thepage}}}}}\@gtempa
   \if@nobreak \ifvmode\nobreak\fi\fi\fi\@esphack}
        \gdef\@eqnlabel{#1}}
\def\@eqnlabel{}
\def\@vacuum{}
\def\draftmarginnote#1{\marginpar{\raggedright\scriptsize\tt#1}}
\def\draft{\oddsidemargin 0.0truein
        \def\@oddfoot{\sl preliminary draft \hfil
        \rm\thepage\hfil\sl\today\quad\militarytime}
        \let\@evenfoot\@oddfoot \overfullrule 3pt
        \let\label=\draftlabel
        \let\marginnote=\draftmarginnote
   \def\@eqnnum{(\theequation)\rlap{\kern\marginparsep\tt\@eqnlabel}%
\global\let\@eqnlabel\@vacuum}  }
\catcode`@=12

\documentstyle[epsf, 12pt]{article}
\begin{document}

\thispagestyle{empty}
\begin{flushright}
RU05-02-B
\end{flushright}

\bigskip\bigskip
\begin{center}
\Large{\bf THE MEISSNER EFFECT IN A TWO FLAVOR LOFF 
COLOR SUPERCONDUCTOR}
\end{center}

\vskip 1.0truecm

\centerline{\bf Ioannis
Giannakis${}^{a}$ and Hai-Cang Ren${}^{a, b}$
\footnote{giannak@summit.rockefeller.edu,
ren@summit.rockefeller.edu}}
\vskip5mm

\centerline{${(a)}${\it Physics Department, Rockefeller University, }}
\centerline{\it New York, NY 10021, U. S. A.}
\centerline{${(b)}${\it Institute of Particle Physics,
Central China Normal University,}}
\centerline{\it Wuhan, 430079, China}

\vskip5mm

\bigskip \nopagebreak \begin{abstract}
\noindent
We calculate the magnetic polarization tensor of 
the photon and of the gluons in a two flavor color
superconductor with a LOFF pairing
that consists of a single plane wave. We show that at zero temperature
and within the range of the values of the Fermi sea displacement
that favors the LOFF state, all the
eigenvalues of the magnetic polarization tensor are non-negative.
Therefore the chromomagnetic instabilities pertaining to a gapless
color superconductor disappear.

\end{abstract}

\newpage\setcounter{page}1

\vfill\vfill\break

\section{Introduction}

The LOFF ( Larkin-Ovchinnikov-Fudde-Ferrell ) state \cite{loff},
\cite{takada} 
was introduced as a candidate superconducting state
of an alloy containing paramagnetic impurities with a ferromagnetic
spin alignment. The presence of a nonzero expectation value
for the spins of the impurities and its corresponding coupling
to the electrons gives rise to a displacement of the Fermi surfaces
of the pairing electrons since they have opposite spins.
The characteristic feature of this state is that the corresponding
order parameter is not constant but depends on the coordinates.
Though the
experimental evidence for the LOFF state is still inconclusive, 
the subject has experienced a revived interest due to its relevance 
to the physics of color superconductors and compact
stars \cite{reviews}.

The interaction between two quarks that
renders quark matter at ultra high baryon
density unstable to the formation of Cooper pairs and leads to
the emergence of a color superconducting state, results from 
the perturbative one-gluon exchange in the color antisymmetric 
diquark channel.
All quark flavors, $u$, $d$ and $s$ can be treated as
massless since the Fermi energies of the quarks are much higher than
$\Lambda_{QCD}$. In contrast at moderately high baryon 
densities, say the density inside a compact star,
 where the Fermi energies of the quarks 
are only few hundred MeV,
instanton effects are expected to dominate the interaction responsible for
the pairing of quarks \cite{rapp}, again in the color antisymmetric
channel.
Since the main pairing channel is between quarks of different flavors,
factors such as the large mass of 
the $s$ quark and the different magnitudes of electric charges of the
$u$ and $d$ quarks together with the 
color/electric neutrality condition induce a considerable displacement 
between the Fermi momenta of the
pairing quarks. This makes the search for a realistic 
color superconductivity phase challenging.

The LOFF state is one of several states that were
proposed as a possible ground state for quark matter at moderately
high baryon density \cite{rajagopal}, \cite{GLR}. Others include
the homogeneous gapless states (g2SC or gCFL ) \cite{shovkovy}, 
\cite{alford}, \cite{phase} and the heterogeneous mixture of BCS
states without Fermi sea displacement and normal phases \cite{bedaque}. 
The gapless state is the analog of the Sarma state
\cite{sarma} for quark matter and a potential ground
state since the Sarma instability can be removed by imposing the 
charge neutrality constraint \cite{shovkovy}, \cite{wilczek}. 
But a subsequent calculation revealed that
the squares of the Meissner masses of the gluons were all negative
signaling a chromomagnetic instability in the presence of 
the gauge field \cite{huang}, \cite{casalbuoni}, \cite{giannak},
\cite{AlfordWang}. This instability represents a major
problem that needs to be resolved before we would be able to
identify the ground state of the quark matter.

The mixed state on the other hand is free from
chromomagnetic instabilities but it has the drawback that it 
is difficult to make a quantitative comparison of
its free energy with one of the homogeneous CSC state. 

In a previous paper \cite{giannak}, we explored the relationship 
between the chromomagnetic instability and the LOFF state.
For a g2SC, we showed 
explicitly that the chromomagnetic instability associated 
with the 8th gluon implies
the existence of a LOFF state with a lower free energy and 
a positive Meissner mass square for the 8th gluon excitation.
But the possibility remains that the
chromomagnetic instability will emerge when we calculate the
Meissner masses of the gluons that are associated with
the 4-7th generators of the color group. 
In this paper we shall continue this 
project by calculating the Meissner masses of all the gluons 
of a LOFF state. 
Our result shows that the chromomagnetic instability
is completely removed at zero temperature
and in the LOFF window where the LOFF state is
energetically favored for a given Fermi momentum displacement. 

This paper is organized as follows: The free energy function of
a two flavor LOFF
CSC will be derived in section II. A discussion
of its properties will also be included. In
section III, the Meissner masses of all the gluonic
excitations will be calculated. The analytic proof of the absence 
of the chromomagnetic instabilities for a small gap parameter
$\Delta$, at $T=0$ will be presented in section IV and 
the numerical results for an arbitrary $\Delta$ will be discussed 
in section V, where we extend the chromagnetic instability free 
region to the entire LOFF window at $T=0$. Finally, in
section VI will summarize our results. Some technical details will 
be presented in the appendices.
Unless explicitly stated, we shall work exclusively with the single 
plane wave ansatz for which analytical investigation can be carried out
quite far.
We shall also designate the symbol
2SC-LOFF for a two flavor LOFF color superconducting state.  

\section{The Free Energy Function of a 2SC-LOFF}

We consider the pairing of two quark flavors, $u$ and $d$.
The NJL effective Lagrangian in the massless limit is given
by \cite{hzc},
\begin{eqnarray}
{\cal L} = -\bar\psi\gamma_\mu\frac{\partial\psi}{\partial x_\mu}
+\bar\psi\gamma_4\mu\psi
&+&G_S[(\bar\psi\psi)^2+(\bar\psi\vec\tau\psi)^2]\nonumber \\
&+&G_D(\bar\psi_C\gamma_5\epsilon^c\tau_2\psi)
(\bar\psi\gamma_5\epsilon^c\tau_2\psi_C)
\label{NJL}
\end{eqnarray}
where $\psi$ represents the quark fields and $\psi_C=C\tilde{\bar\psi}$
is its charge conjugate with $C=i\gamma_2\gamma_4$. All gamma 
matrices are hermitian,
$(\epsilon^c)^{mn}=\epsilon^{cmn}$ is 
a $3\times 3$ matrix acting on the red, green and blue color indices and 
the Pauli matrices $\vec\tau$ act on the $u$ and $d$ flavor (isospin) 
indices.
The chemical potential is also a matrix in color-flavor space, i.e.
\begin{equation}
\mu = \bar\mu-\delta\tau_3+\delta^\prime\lambda_8
\end{equation}
with $\lambda_l$ being the $l$th Gell-Mann matrix.

The most general expression of the
order parameter of 2SC-LOFF is a superposition of plane
waves of diquark condensate, i.e.  
\begin{equation}
<\bar\psi(\vec r)\gamma_5\lambda_2\tau_2\psi_C(\vec r)>
=\sum_{\vec q}\Phi_{\vec q} e^{2i\vec q\cdot\vec r},
\label{order_general}
\end{equation}
In this article, we shall be mainly working with the type of the
LOFF state that 
contains only one plane wave, more specifically the form of the
order parameter will be given by 
\begin{equation}
<\bar\psi(\vec r)\gamma_5\lambda_2\tau_2\psi_C(\vec r)>
=\Phi e^{2i\vec q\cdot\vec r}.
\label{order}
\end{equation}
The condensate eq. (\ref{order}) is invariant under a 
translation accompanied by a phase shift. Since the free energy depends
on the magnitude of the condensate only, it is translationally 
invariant. But this is not clearly the 
case with the general form of the 2SC-LOFF condensate eq. (\ref{order_general}).

Let's introduce a transformed quark field, 
\begin{equation} 
\chi(\vec r)=e^{-i\vec q\cdot\vec r}\psi(\vec r).
\label{loff}
\end{equation}
The NJL Lagrangian in terms of the new quark field reads
\begin{eqnarray}
{\cal L} = -\bar\chi\gamma_\mu\frac{\partial\chi}{\partial x_\mu}
-i\bar\chi\vec\gamma\cdot\vec q\chi+\bar\chi\gamma_4\mu\chi
&+& G_S[(\bar\chi\chi)^2+(\bar\chi\vec\tau\chi)^2]\nonumber \\
&+&G_{D}(\bar\chi_C\gamma_5\epsilon^c\tau_2\chi)
(\bar\chi\gamma_5\epsilon^c\tau_2\chi_C)
\label{nambu},
\end{eqnarray}
while the condensate (\ref{order}) takes the form
\begin{equation}
<\bar\chi(\vec r)\gamma_5\lambda_2\tau_2\chi_C(\vec r)>=\Phi,
\label{order_chi}
\end{equation}

By ignoring the chiral condensate and expanding the NJL
Lagrangian to linear order in the fluctuation
\begin{equation}
\bar\psi\gamma_5\lambda_2\tau_2\psi_C-\Phi,
\label{order1}
\end{equation}
we derive the following mean field expression for the NJL Lagrangian
\begin{eqnarray}
{\cal L}_{MF}=-\frac{\Delta^2}{4G_D}
-\bar\chi\gamma_\mu\frac{\partial\chi}{\partial x_\mu}
-i\bar\chi\vec\gamma\cdot\vec q\chi+\bar\chi\gamma_4\mu\chi
+\frac{1}{2}\Delta(-\bar\chi_C\gamma_5\lambda_2\tau_2\chi
+\bar\chi\gamma_5\lambda_2\tau_2\chi_C),
\label{NJL_MF}
\end{eqnarray}
where the gap parameter is given by $\Delta=2G_D\Phi>0$.
The free energy ${\cal F}$ is given by 
the Euclidean path integral
\begin{equation}
{\cal F}=-k_BT\ln\Big[\int[d\psi d\bar\psi]
\exp\Big(\int d^4x{\cal L}_{MF}\Big)\Big]
\label{free_energy}
\end{equation}
with $0<x_4<(k_BT)^{-1}$ and will be minimized at equilibrium.
We find it convenient to work with the difference of the free energy 
density between the superconducting and normal phases in a
grand canonical ensemble
at the same 
$\delta$ and $\delta^\prime$,
\begin{equation}
\Gamma=\frac{1}{\Omega}({\cal F}-{\cal F}\vert_{\Delta=0}),
\end{equation}
where $\Omega$ is the volume of the system.
The equilibrium conditions read
\begin{equation}
\frac{\partial\Gamma}{\partial\Delta^2}=0, \qquad 
\frac{\partial\Gamma}{\partial q}=0,
\label{thermal}
\end{equation}
and
\begin{equation}
\frac{\partial^2\Gamma}{(\partial\Delta^2)^2}>0, \qquad
\left|\matrix{\frac{\partial^2\Gamma}{(\partial\Delta^2)^2}&
\frac{\partial^2\Gamma}{\partial\Delta^2\partial q}\cr
\frac{\partial^2\Gamma}{\partial\Delta^2\partial q}&
\frac{\partial^2\Gamma}{\partial q^2}\cr}\right|>0.
\label{stability}
\end{equation}
It follows then that 
\begin{equation}
\frac{\partial^2\Gamma}{\partial q^2}>0
\end{equation}
at the minimum.

In terms of the Nambu-Gorkov representation of the quark field, 
\begin{equation}
\Psi=\left(\matrix{\chi\cr \tau_2\chi_C\cr}\right), \qquad \bar\Psi
=(\bar\chi,\bar\chi_C\tau_2)
\label{NGbasis}
\end{equation}
and its Fourier transform, 
\begin{equation}
\Psi(\vec r)=\sqrt{k_BT}{\sum_\nu}\int\frac{d^3\vec p}
{(2\pi)^3}e^{iPx}\Psi_P
\end{equation}
we find that
\begin{equation}
\int d^4x{\cal L}_{MF} = \Omega\Big[-\frac{\Delta^2}{4G_D}
+\frac{1}{2}\sum_\nu\int\frac{d^3\vec p}{(2\pi)^3}\bar\Psi_P
{\cal S}^{-1}(P)\Psi_P\Big]
\end{equation}
where ${\cal S}^{-1}(P)$ is the momentum representation of 
the inverse quark propagator
\begin{eqnarray}
{\cal S}^{-1}(P) &=& \left( \begin{array}{cc} 
/\kern-7pt P-i\vec\gamma\cdot\vec q+\bar\mu\gamma_4
-\delta\gamma_4\tau_3+\delta^\prime\gamma_4\lambda_8 & \Delta\gamma_5\lambda_2 \\ 
-\Delta\gamma_5\lambda_2 & /\kern-7pt P+i\vec\gamma\cdot\vec q
-\bar\mu\gamma_4-\delta\gamma_4\tau_3-\delta^\prime\gamma_4\lambda_8
\end{array} \right)\nonumber \\
&=& /\kern-7pt P-\delta\gamma_4\tau_3
+(-i\vec\gamma\cdot\vec q+\bar\mu\gamma_4
+\delta^\prime\gamma_4\lambda_8)\rho_3+i\Delta\gamma_5\lambda_2\rho_2.
\label{prop_gap}
\end{eqnarray}
$/\kern-7pt P=-i\gamma_\mu P_\mu$, $P=(\vec p,-\nu)$ with
$\nu=(2n+1)\pi k_BT$ being the Matsubara frequency, and we have 
introduced the Pauli matrices with respect to NG blocks, i.e.
\begin{equation}
\rho_1=\left(\matrix{0&I\cr I&0\cr}\right) \qquad 
\rho_2=\left(\matrix{0&-iI\cr iI&0\cr}\right) \qquad
\rho_3=\left(\matrix{I&0\cr 0&-I\cr}\right).
\end{equation}
The product of Dirac, color, flavor and NG matrices is a 
direct product and the total dimensionality of ${\cal S}^{-1}(P)$
is $4({\rm Dirac})\times 3({\rm color})\times 2({\rm flavor})
\times2({\rm NG})=48$.

Carrying out the path integration in Eq. (\ref{free_energy}), we find
\begin{equation}
\Gamma=\frac{\Delta^2}{4G_D}-\frac{k_BT}{2}\sum_\nu\int\frac{d^3\vec p}{(2\pi)^3}
\ln\frac{{\rm det}{\cal S}^{-1}(P)}{{\rm det}S^{-1}(P)},
\label{gamma}
\end{equation}
where $S(P)\equiv{\cal S}(P)\vert_{\Delta=0}$ is the free quark propagator. 
 
The typical energy scale where the
pairing occurs is characterized by the zero temperature gap parameter 
$\Delta_0$ in the absence of a displacement, i.e. $\delta=\delta^\prime=0$. 
At weak coupling, $\Delta_0<<\bar\mu$ and the pairing interaction 
extends over a
width of $\omega_0$ on both sides of the Fermi level with $\Delta_0
<<\omega_0<<\bar\mu$. Furthermore the displacement $\delta^\prime$
that is induced by the 
condensate, is of higher order and can be neglected. As a
result we have the only displacement parameter $\delta\sim\Delta_0$.
Under this approximation, the quark propagator takes the approximate form
\begin{equation}
{\cal S}(P)\simeq\frac{(-i\vec\gamma\cdot\hat p+\rho_3\gamma_4)}{2}
\frac{[|\vec p-\rho_3\vec q|-\bar\mu+\rho_3(i\nu-\delta\tau_3)
+\Delta\lambda_2\gamma_5\gamma_4]}{(p-\bar\mu)^2+
(i\nu-\delta\tau_3-\hat p\cdot\vec q)^2+\Delta^2}
\label{approx_S}
\end{equation}
while the determinant in eq. (\ref{gamma}) is given by
\begin{equation}
{\rm det}{\cal S}^{-1}(P) = (2\bar\mu)^{24}\prod_{s,\lambda}[(p-\bar\mu)^2-
(i\nu+s\delta-\hat p\cdot\vec q)^2+(\lambda\Delta)^2]^2
\label{deter_approx}
\end{equation}
where $s=\pm 1$ and $\lambda=\pm 1, 0$ refer to the eigenvalues of $\tau_3$ 
and $\lambda_2$ respectively. A more detailed derivation
of eqs. (\ref{approx_S}) 
and (\ref{deter_approx}) can be found in 
appendix A. It follows from the analytic continuation of (\ref{approx_S})
to the real energy, $i\nu\to p_0$, that the low-lying excitation 
spectrum reads
\begin{equation}
E_{\vec p}^{\pm}=\sqrt{(p-\bar\mu)^2+\Delta^2}\pm\delta-\hat p\cdot\vec q
\label{spectrum}
\end{equation}
for red and green colors and $E_{\vec p}=|p-\bar\mu|
\pm\delta-\hat p\cdot\vec q$ for blue color. 
The excitations (\ref{spectrum}) are fully gapped if 
$\Delta>\delta+q$. $E_{\vec p}^-$ becomes gapless when 
$|\delta-q|<\Delta<\delta+q$ and both $E_{\vec p}^{\pm}$
become gapless when $q>\delta$ and $q-\delta<\Delta<q+\delta$
\cite{takada}. 

Substituting (\ref{deter_approx}) into (\ref{gamma}) we obtain that 
\begin{equation}
\Gamma=\frac{\Delta^2}{4G_D}-\frac{\bar\mu^2k_BT}{\pi^2}\sum_{|\nu|<\omega_0}
\int_{-1}^1 d\cos\theta\int_{-\infty}^\infty
d\xi\ln\frac{\xi^2-z^2(\nu,\theta,\Delta)}{\xi^2-z^2(\nu,\theta, 0)}
\end{equation}
with $\xi=p-\bar\mu$ and $z(\nu,\theta,\Delta)
=\sqrt{(i\nu+\delta-q\cos\theta)^2-\Delta^2}$.
The contributions from the eigenvalues of $\tau_3$ with opposite
sign are equal if we
reverse the signs
of the Matsubara frequency and simultaneously perform the
transformation $\theta\to\pi-\theta$. 
This will also be the case with the calculation of the
Meissner masses in the next section.

We shall employ the phase conventions that are determined by
\begin{equation} 
{\rm Im}z(\nu,\theta,\Delta)>0 \hbox{ for $\nu>0$ },
\label{phase}
\end{equation}
with $\zeta(-\nu, \theta, \Delta)=\zeta^{\star}(\nu, \theta, \Delta)$.
Let's now proceed and perform the $\xi$ integral by closing 
the contour around one of the branch cuts of the logarithm.
As a result we obtain the expression
\begin{equation}
\Gamma = \frac{\Delta^2}{4G_D}-\frac{4\bar\mu^2k_BT}{\pi}{\rm Im}
\int_{-1}^1d\cos\theta\sum_{0<\nu<\omega_0}
[z(\nu,\theta,\Delta)-z(\nu,\theta,0)].
\end{equation}

At zero temperature, $T=0$, we have
\begin{equation}
k_BT\sum_{0<\nu<\omega_0}(...)=\int_{0^+}^{\omega_0}\frac{d\nu}{2\pi}(...)
\end{equation}
and
\begin{eqnarray}
\Gamma &=& \frac{\Delta^2}{4G_D}-\frac{2\bar\mu^2}{\pi^2}\Delta^2
\Big(\ln\frac{2\omega_0}{\Delta}+\frac{1}{2}\Big) \nonumber \\
& + & \frac{\bar\mu^2}{\pi^2}
\int_{-1}^1d\cos\theta\Big[\theta(|\delta_\theta|-\Delta)
(\Delta^2\ln\frac{|\delta_\theta|+\sqrt{\delta_\theta^2
-\Delta^2}}{\Delta}-|\delta_\theta|\sqrt{\delta_{\theta}^2-\Delta^2})
+\delta_{\theta}^2\Big]
\label{egw}
\end{eqnarray}
where $\delta_\theta=(1-\rho\cos\theta)\delta$ with $\rho=q/\delta$, 
being the normalized diquark momentum.
The limit of the angular
integral depends on the gapless region of the excitation spectrum
(\ref{spectrum}) and
will be discussed in detail in the appendix B. In terms of the gap parameter 
$\Delta_0$ at $\delta=0$, given by the standard BCS gap equation
\begin{equation}
\frac{1}{4G_D}=\frac{2\bar \mu^2}{\pi^2}\ln\frac{2\omega_0}{\Delta_0},
\end{equation}
we have
\begin{eqnarray}
\Gamma=\frac{2{\bar\mu}^2}{\pi^2}\lbrace\Delta^2\Big
(\ln\frac{\Delta}{\Delta_0}
& -& \frac{1}{2}\Big)+\frac{(\rho+1)^3}{4\rho}\delta^2
\Big[(1-x_1^2)\ln\frac{1+x_1}{1-x_1}
+\frac{2}{3}(2x_1^3-3x_1+1)\Big]\nonumber \\
&+&\frac{(\rho-1)^3}{4\rho}\delta^2\Big[(1-x_2^2)
\ln\frac{1+x_2}{1-x_2}+\frac{2}{3}(2x_2^3-3x_2+1)\Big]\rbrace,
\label{freeenergy}
\end{eqnarray}
where $x_1$ and $x_2$ are the same dimensionless parameters introduced in 
Ref. \cite{takada}, i.e.
\begin{equation}
x_1=\theta\Big(1-\frac{\Delta}{(\rho+1)\delta}\Big)
\sqrt{1-\frac{\Delta^2}{(\rho+1)^2\delta^2}}
\label{malakia}
\end{equation}
and 
\begin{equation}
x_2=\theta\Big(1-\frac{\Delta}{|\rho-1|\delta}\Big)
\sqrt{1-\frac{\Delta^2}{(\rho-1)^2\delta^2}}.
\label{chi2}
\end{equation}
Differentiating eq. (\ref{freeenergy}) with respect to $\Delta^2$ and 
$\rho$, we derive the equilibrium equations
\begin{equation}
\frac{\partial\Gamma}{\partial\Delta^2}=\frac{2\bar\mu^2}{\pi^2}
\Big[\ln\frac{\Delta}{\Delta_0}
-\frac{\rho+1}{4\rho}\Big(\ln\frac{1-x_1}{1+x_1}+2x_1\Big)
-\frac{\rho-1}{4\rho}\Big(\ln\frac{1-x_2}{1+x_2}+2x_2\Big)\Big]=0
\label{equil_1}
\end{equation}
and 
\begin{eqnarray}
\frac{\partial\Gamma}{\partial\rho}=\frac{4\bar\mu^2\delta^2}{3\pi^2}
\rho\{1&-&\frac{(\rho+1)^2}{8\rho^3}\Big[3(1-x_1^2)\ln\frac{1+x_1}{1-x_1}
+4(\rho+1)x_1^3-6x_1\Big]\nonumber \\
&-&\frac{(\rho-1)^2}{8\rho^3}\Big[-3(1-x_2^2)\ln\frac{1+x_2}{1-x_2}
+4(\rho-1)x_2^3+6x_2\Big]\}=0.
\label{equil_2}
\end{eqnarray} 
If we subsitute into (\ref{freeenergy}) 
the solution to the gap equation (\ref{equil_1}), 
we find
\begin{equation}
\Gamma = \frac{2\bar\mu^2}{\pi^2}\{\delta^2+\frac{1}{3}\rho^2\delta^2
-\frac{1}{2}\Delta^2-\frac{1}{6\rho}
[(\rho+1)^3x_1^3+(\rho-1)^3x_2^3]\delta^2\}.
\label{free_min}
\end{equation}

Furthermore, the second order derivatives provide us with the expressions
\begin{equation}
\frac{\partial^2\Gamma}{(\partial\Delta^2)^2}
=\frac{\bar\mu^2}{2\pi^2\rho\delta^2}\Big[\frac{1}{(\rho+1)(1+x_1)}
+\frac{1}{(\rho-1)(1+x_2)}\Big],
\end{equation}
\begin{eqnarray}
\frac{\partial^2\Gamma}{\partial\rho^2} &=& \frac{4\bar\mu^2\delta^2}{3\pi^2}
\Big[1+\frac{3(\rho+1)^2}{4\rho^3}(1-x_1^2)\ln\frac{1+x_1}{1-x_1}
-\frac{3(\rho-1)^2}{4\rho^3}(1-x_2^2)\ln\frac{1+x_2}{1-x_2}\nonumber \\
&+& \frac{(\rho+1)^3}{\rho^3}x_1^3-\frac{3(\rho+1)(\rho^2+\rho+1)}{2\rho^3}x_1
+\frac{(\rho-1)^3}{\rho^3}x_2^3-\frac{3(\rho-1)(\rho^2-\rho+1)}{2\rho^3}x_2
\Big]
\label{exp_D}
\end{eqnarray}
and
\begin{equation}
\frac{\partial^2\Gamma}{\partial\Delta^2\partial\rho}
=\frac{\bar\mu^2}{2\pi^2\rho^2}\Big[\ln\frac{1-x_1}{1+x_1}
-\ln\frac{1-x_2}{1+x_2}+2(\rho+1)x_1+2(\rho-1)x_2\Big].
\end{equation}

Our expression of the free energy function eq. (\ref{free_min}) 
coincides with the expression derived in \cite{takada}
if we perform the following modifications:
replace
the Fermi velocity in their formula with one and 
multiply it by a factor of four. This factor accounts 
for the four pairing configurations, $(r,u_R)-(g,d_R)$, $(r,u_L)-(g,d_L)$,
$(r,d_R)-(g,u_R)$ and $(r,d_L)-(g,u_L)$, per momentum for quarks 
versus one pairing configuration, spin up-spin down,
per momentum for electrons, where the subscripts R and L refer
to the helicities of the pairing quarks.

\section{The Meissner Masses}

In this section we shall consider the coupling of the quark fields 
to external gauge fields.
The mean field NJL Lagrangian is modified
\begin{eqnarray}
{\cal L}& = &-\frac{\Delta^2}{4G_D}+\bar\chi\gamma_\mu\Big(\frac{\partial}
{\partial x_\mu}-igA_\mu^lT_l-ieQA_\mu\Big)\chi
-\bar\chi{\vec\gamma}{\vec q}\chi+\bar\chi\gamma_4\mu\chi\nonumber \\
&+&\Delta(-\bar\chi_C\gamma_5\lambda_2\tau_2\chi
+\bar\chi\gamma_5\lambda_2\tau_2\chi_C).
\label{NJL_gauge}
\end{eqnarray}
$A_\mu^l$ and $A_\mu$ represent the gluon and photon gauge potentials
respectively.
The Lagrangian is invariant under $SU(3)_c\times U(1)_{em}$
transformations. The $SU(3)_c$ generators and the $U(1)_{em}$
charge operator can be written as
\begin{equation}
T_l=\frac{1}{2}\lambda_l, \qquad Q=\frac{1}{6}+\frac{1}{2}\tau_3.
\label{charge}
\end{equation}

The squared Meissner masses are the eigenvalues of the Meissner
tensors, defined by
\begin{equation}
(M^2)_{ij}^{l^\prime l}=\lim_{K\to 0}
\Pi_{ij}^{l^\prime l}(K),
\end{equation}
where $\Pi_{ij}^{l^\prime l}(K)$ is the gluon/photon magnetic 
polarization tensor given by
\begin{eqnarray}
\Pi_{ij}^{l^\prime l}(K) &=& \frac{1}{2}\int_0^{(k_BT)^{-1}}dx_4
\int d^3\vec r
e^{-iK\cdot x}<J_i^{l^\prime}(\vec r,\tau)J_j^l(0,0)>_C\nonumber \\
&=& -\frac{k_BT}{2}\sum_\nu\int\frac{d^3\vec p}{(2\pi)^3}
Tr\Gamma_i^{l^\prime}{\cal S}(P)\Gamma_j^{l}{\cal S}(P+K),
\label{eqkit}
\end{eqnarray}
where $<...>_C$ stands for the connected Green's functions, i.e.
$<AB>_C=<AB>-<A><B>$ and $K=(k\hat z,-\omega)$
is the Euclidean four momentum either of the gluon
or of the photon. The current operator reads
$\vec J^l=\frac{1}{2}\bar\Psi\vec\Gamma^l\Psi$, 
where the quark-gluon/photon vertex with 
respect to the NG basis (\ref{NGbasis}) takes the form
\begin{equation}
\Gamma_\mu^{\prime l} =\left( \begin{array}{cc}
\gamma_\mu {\cal T}_l & 0 \\
 0 & -\gamma_\mu\tau_2\tilde{\cal T}_l\tau_2
\end{array} \right),
\end{equation}
where $l=1,2,...,9$ and the tilde denotes transpose. Here we have 
introduced a new set of generators, ${\cal T}_l$, that mixes
the gluons with photon,
in accordance with the residual symmetry in 
the presence of the condensate (\ref{order})\cite{mixing}, i.e. 
\begin{equation}
{\cal T}_l=T_l
\label{set1}
\end{equation}
for $l=1,...,7$,
\begin{equation}
{\cal T}_8=T_8\cos\theta+\frac{e}{g}Q\sin\theta
\label{set2}
\end{equation}
and
\begin{equation}
{\cal T}_9=-\frac{g}{e}T_8\sin\theta+Q\cos\theta
\label{set3}
\end{equation}
with 
$\cos\theta=\frac{\sqrt{3}g}{\sqrt{3g^2+e^2}}$ and $\sin\theta
=\frac{e}{\sqrt{3g^2+e^2}}$.  
The condensate is invariant under the residual gauge transformations 
$SU(2)\times{\cal U}(1)$, where $SU(2)$ is generated by  
${\cal T}_1$, ${\cal T}_2$, ${\cal T}_3$ and ${\cal U}(1)$
by ${\cal T}_9$. We shall refer to the photon-gluon mixture
associated to ${\cal T}_8$ as the 8th gluon.

Each component of the Meissner tensor is a matrix with respect to the 
generators of the gauge group, eqs.(\ref{set1}), (\ref{set2}) and 
(\ref{set3}).
A simplification follows from the observation that 
the isospin part of the quark-photon vertex is the unit matrix with respect to 
the NG blocks ( $-\tau_2\tilde\tau_3\tau_2=\tau_3$ )
and consequently it contributes to the integrand of the Meissner tensor
a total derivative,
i.e.
\begin{equation}
\int\frac{d^3\vec p}{(2\pi)^3}
Tr\gamma_j\tau_3{\cal S}(P){\cal M}{\cal S}(P)
= i\int\frac{d^3\vec p}{(2\pi)^3}
\frac{\partial}{\partial p_j}Tr\tau_3{\cal M}{\cal S}(P),
\end{equation}
where ${\cal M}$ is an arbitrary matrix. Therefore the isospin term of 
the charge operator (\ref{charge}) can be omitted and we 
have effectively $Q=\frac{1}{6}$ \cite{giannak, AlfordWang}.

Next we shall prove that the Meissner tensor is diagonal with respect 
to the indices $l$ and $l^\prime$, more specifically
that it has the following form
\begin{equation}
(M^2)_{ij}^{11}=(M^2)_{ij}^{22}=(M^2)_{ij}^{33}=(M^2)_{ij}^{99}=0,
\end{equation}
\begin{equation}
(M^2)_{ij}^{44}=(M^2)_{ij}^{55}=(M^2)_{ij}^{66}=(M^2)_{ij}^{77}
\equiv (m^2)_{ij},
\end{equation}
and 
\begin{equation}
(M^2)_{ij}^{88}\equiv (m^{\prime 2})_{ij}.
\end{equation}

The relations above
follow in a straightforward manner from symmetry arguments, in
particular the residual $SU(2)$ symmetry.
We proceed to divide the set of the nine generators into three 
subsets.
Subset I includes the unbroken generators ${\cal T}_1$, 
${\cal T}_2$ and ${\cal T}_3$,
subset II includes the broken generators ${\cal T}_4$, 
${\cal T}_5$, ${\cal T}_6$
and ${\cal T}_7$ while subset III includes the broken ${\cal T}_8$ and the 
unbroken ${\cal T}_9$ generators. Under an $SU(2)$ transformation,
${\cal T}_l\to u{\cal T}_lu^\dagger$,
the generators in the subset I transform as the adjoint representation, 
the generators in the subset II as the fundamental 
one while the generators in the subset III 
remain invariant. Therefore the indices $l^\prime$ 
and $l$ which correspond to generators from different subsets
do not mix in the expression for $(M^2)^{l^\prime l}_{ij}$. The mass 
matrix vanishes within the subset I because of the unbroken gauge symmetry.
As for generators of the subset II, we proceed 
to relabel its members according 
to ${\cal J}_1=T_4$, ${\cal J}_{\bar 1}
=T_5$, ${\cal J}_2=T_6$ and ${\cal J}_{\bar 2}=T_7$. Introducing 
\begin{equation}
\phi_1= \left( \begin{array}{cc} 1 \\ 0 \end{array} \right) \hbox{\\\\\\\\}
\qquad \phi_2= \left( \begin{array}{cc} 0 \\ 1\end{array} \right),
\end{equation}
we have 
\begin{equation}
{\cal J}_\alpha=\frac{1}{2}\left( \begin{array}{cc} 0 
& \phi_\alpha \\ \phi_\alpha^\dagger & 0 \end{array} \right),
\end{equation}
and similar expressions for ${\cal J}_{\bar\alpha}$ with 
$\phi_{\bar\alpha}=-i\phi_\alpha$, where 
the block decomposition is with respect to color indices.
Since ${\cal J}_\alpha$ is symmetric and
${\cal J}_{\bar\alpha}$ is antisymmetric, there is
no mixing between ${\cal J}_\alpha$ and
${\cal J}_{\bar\beta}$, following from the properties 
of the matrix $C=i\gamma_2\gamma_4$, i.e.
$C\tilde\gamma_\mu C^{-1}=-\gamma_\mu$ and
\begin{equation} 
\rho_3C\tilde{\cal S}(P)(\rho_3C)^{-1}=-{\cal S}(P),
\label{TPgamma5}
\end{equation}
a consequence of the symmetry under a combined transformation 
of a space inversion, a time reversal operation and a $\gamma_5$ 
multiplication (see appendix C for details).
The identity of the matrix elements $(M^{2})^{\alpha\beta}$ 
and $(M^{2})^{\bar\alpha\bar\beta}$ follows from the equivalence 
of the doublet representation supported by ${\cal T}_\alpha$ and 
the anti-doublet one by ${\cal T}_{\bar\alpha}$. 
Finally in the subset III, the color matrices
never mix the red and green components with the 
blue ones. The latter do not carry a condensate. It follows 
then that we need only to work 
within the red-green subspace, in which ${\cal T}_9$ does not 
contribute and ${\cal T}_8$ 
may be replaced by a factor $\frac{1}{6g}\sqrt{3g^2+e^2}$.

The above argument on the group theoretic structure of the Meissner 
tensor leads also to its relation with the momentum susceptibility
introduced in Ref. \cite{giannak}, 
\begin{equation}
(m^\prime)^2_{ij}=\frac{1}{12}\Big(g^2+\frac{e^2}{3}\Big)
\frac{\partial^2\Gamma}{\partial q_i\partial q_j}.
\label{susc}
\end{equation}

Because of the single plane wave ansatz, eq. (\ref{order}), each of 
the nontrivial tensors $(m^2)_{ij}$ and $(m^{\prime 2})_{ij}$ 
can further be decomposed 
into its transverse and logitudinal components, i.e.
\begin{equation}
(m^2)_{ij}=A(\delta_{ij}-\hat q_i\hat q_j)+B\hat q_i\hat q_j
\end{equation}
\begin{equation}
(m^{\prime 2})_{ij}=C(\delta_{ij}-\hat q_i\hat q_j)+D\hat q_i\hat q_j.
\end{equation}
In particular, eq. (\ref{susc}) implies that
\begin{equation}
C=\frac{g^2}{12\rho\delta^2}\Big(1+\frac{e^2}{3g^2}\Big)
\frac{\partial\Gamma}{\partial\rho}=
\frac{g^2}{6\delta^2}\Big(1+\frac{e^2}{3g^2}\Big)
\frac{\partial\Gamma}{\partial\rho^2}
\label{teri}
\end{equation}
and
\begin{equation}
D=\frac{g^2}{12\delta^2}\Big(1+\frac{e^2}{3g^2}\Big)
\frac{\partial^2\Gamma}{\partial\rho^2}.
\label{eteri}
\end{equation}

It follows from eqs. (\ref{teri}) and (\ref{equil_2})
at $\rho\neq 0$ that the Meissner tensor of the 8th gluon
is longitudinal at the LOFF equilibrium. So is, the
Meissner tensor of an electronic LOFF superconductor. This
conclusion of ours disagrees with that of \cite{takada},
where a nonzero transverse Meissner mass was reported.
The above property does not contradict with the $\delta_{ij}$
structure of the Meissner tensor of a homogeneous superconductor
since it corresponds to the trivial solution,
$\rho=0$ of eq. (\ref{equil_2}), where $\frac{\partial\Gamma}
{\partial\rho^2}\neq 0$.

In order to  calculate the quantities $A$, $B$, $C$ and $D$ 
we follow the same 
procedure that we used to calculate
the free energy in the previous section. In order to improve 
the UV convergence, we employ the technique of \cite{book} and subtract from 
the polarization tensor of the superconducting phase, (\ref{eqkit}), 
the correponding expression of the normal phase with the same 
$\bar\mu$ and $\delta$, which is zero at $K=0$, and write
\begin{equation}
\Pi_{ij}^{l^\prime l}(0)=-\frac{k_BT}{2}
\sum_\nu\int\frac{d^3\vec p}{(2\pi)^3}
Tr[\Gamma_i^{l^\prime}{\cal S}(P)\Gamma_j^{l}{\cal S}(P)
-\Gamma_i^{l^\prime}S(P)\Gamma_j^{l}S(P)],
\label{eqkit1}
\end{equation}
where $S(P)={\cal S}(P)\mid_{\Delta=0}$. Then the
weak coupling approximation
of the propagator, eq.(\ref{approx_S}), is 
inserted. We take the $z$-axis to be parallel to $\vec q$ and carry out 
the integration over the azimuthal angle and the
magnitude of $\vec p$. We are
then left with the integration over the meridian 
angle and the summation over 
the Matsubara energies. Finally, we project out 
the transverse and longitudinal
components and obtain the expressions
\begin{equation}
A = \frac{3}{4}\int_{-1}^1d\cos\theta\sin^2\theta F(\theta,\Delta),
\end{equation}
\begin{equation}
B = \frac{3}{2}\int_{-1}^1d\cos\theta\cos^2\theta F(\theta,\Delta),
\end{equation}
\begin{equation}
C = \frac{3}{4}\int_{-1}^1d\cos\theta\sin^2\theta G(\theta,\Delta),
\end{equation}
and
\begin{equation}
D = \frac{3}{2}\int_{-1}^1d\cos\theta\cos^2\theta G(\theta,\Delta),
\end{equation}
where
\begin{equation}
F(\theta,\Delta)=\frac{2g^2\bar\mu^2k_BT}{3\pi}{\rm Im}\sum_{\nu>0}
\frac{1}{z(\nu,\theta,\Delta)}\frac{i\nu+\delta_\theta-z(\nu,\theta,\Delta)}
{i\nu+\delta_\theta+z(\nu,\theta,\Delta)}
\end{equation}
and
\begin{equation}
G(\theta,\Delta)=\frac{2g^2\bar\mu^2\Delta^2k_BT}{9\pi}
\Big(1+\frac{e^2}{3g^2}\Big){\rm Im}\sum_{\nu>0}
\frac{1}{z^3(\nu,\theta,\Delta)}.
\end{equation}

At $T=0$, the summation over $\nu$ becomes an integral and we find
\begin{equation}
F(\theta,\Delta)=\frac{g^2\bar\mu^2}{3\pi^2}\Big[\frac{1}{2}-\frac{\delta_\theta^2}{\Delta^2}
+\theta(|\delta_\theta|-\Delta)\frac{\delta_\theta^2}{\Delta^2}
\sqrt{1-\frac{\Delta^2}{\delta_\theta^2}}\Big]
\label{HS4_7}
\end{equation}
and
\begin{equation}
G(\theta,\Delta)=\frac{g^2\bar\mu^2}{9\pi^2}\Big(1+\frac{e^2}{3g^2}\Big)
\Big[1-\frac{\theta(|\delta_\theta|-\Delta)}
{\sqrt{1-\frac{\Delta^2}{\delta_\theta^2}}}\Big].
\label{HS8}
\end{equation}

We recognize that eq. (\ref{HS4_7}) and (\ref{HS8}) are precisely 
the expressions that Huang and Shovkovy
obtained for the Meissner masses \cite{huang} when their
parameter $\delta\mu$ is
replaced by an angle-dependent one, $|\delta_\theta|$. By carrying out 
the angular integration
as it is described in detail in the appendix B, we find 
\begin{eqnarray}
A &=& \frac{g^2\bar\mu^2}{6\pi^2}\Big[1-\frac{2\delta^2}{\Delta^2}
\Big(1+\frac{1}{5}\rho^2\Big)
-\frac{3\Delta^2}{16\rho^3\delta^2}\Big(\ln\frac{1+x_1}{1-x_1}
-\ln\frac{1+x_2}{1-x_2}\Big)\\ \nonumber
&+& \frac{\delta^2}{5\rho^3\Delta^2}(\rho+1)^5x_1^5
-\frac{\delta^2}{8\rho^3\Delta^2}(\rho+1)^4x_1(5x_1^2-3)\\ \nonumber
&+& \frac{\delta^2}{5\rho^3\Delta^2}(\rho-1)^5x_2^5
+\frac{\delta^2}{8\rho^3\Delta^2}(\rho-1)^4x_2(5x_2^2-3)\Big],
\end{eqnarray}
\begin{eqnarray}
B &=& \frac{g^2\bar\mu^2}{6\pi^2}\Big[1-\frac{2\delta^2}{\Delta^2}
\Big(1+\frac{3}{5}\rho^2\Big)+\frac{3\Delta^2}{8\rho^3\delta^2}
\Big(\ln\frac{1+x_1}{1-x_1}-\ln\frac{1+x_2}{1-x_2}\Big)\\ \nonumber
&+& \frac{\delta^2}{\rho\Delta^2}(\rho+1)^3x_1^3
-\frac{2\delta^2}{5\rho^3\Delta^2}(\rho+1)^5x_1^5
+\frac{\delta^2}{4\rho^3\Delta^2}(\rho+1)^4x_1(5x_1^2-3)\\ \nonumber
&+& \frac{\delta^2}{\rho\Delta^2}(\rho-1)^3x_2^3
-\frac{2\delta^2}{5\rho^3\Delta^2}(\rho-1)^5x_2^5
-\frac{\delta^2}{4\rho^3\Delta^2}(\rho-1)^4x_2(5x_2^2-3)\Big],
\end{eqnarray}
\begin{eqnarray}
C = \frac{g^2\bar\mu^2}{9\pi^2}\Big(1+\frac{e^2}{3g^2}\Big)
\lbrace1&-&\frac{(\rho+1)^2}{8\rho^3}\Big[3(1-x_1^2)\ln\frac{1+x_1}{1-x_1}
+4(\rho+1)x_1^3-6x_1\Big]\nonumber \\
&-&\frac{(\rho-1)^2}{8\rho^3}\Big[-3(1-x_2^2)\ln\frac{1+x_2}{1-x_2}
+4(\rho-1)x_2^3+6x_2\Big]\rbrace
\label{expC}
\end{eqnarray}
and
\begin{eqnarray}
D = \frac{g^2\bar\mu^2}{9\pi^2}\Big(1+\frac{e^2}{3g^2}\Big)
\{1 &+& \frac{\rho+1}{4\rho^3}\Big[3(\rho+1)(1-x_1^2)\ln\frac{1+x_1}{1-x_1}
\nonumber \\
&+& 4(\rho+1)^2x_1^3-6(\rho^2+\rho+1)x_1\Big]\nonumber \\
&+& \frac{\rho-1}{4\rho^3}\Big[-3(\rho-1)(1-x_2^2)\ln\frac{1+x_2}{1-x_2}
\nonumber \\
&+& 4(\rho-1)^2x_2^3-6(\rho^2-\rho+1)x_2\Big]\}.
\label{expD}
\end{eqnarray}

Comparing (\ref{expC}) and (\ref{expD}) with (\ref{equil_2})
and (\ref{exp_D}), we confirm
explicitly the relation between the Meissner tensor of
the 8th gluon and
the momentum susceptibility, eq. (\ref{susc}).

\section{The Meissner Effect for a Small Gap Parameter}

In the last two sections we
derived the expressions for the free energy and the Meissner
masses of the gluons for a 2SC-LOFF
of arbitrary $\Delta$, $\delta$ and $\rho$. In this section we shall
proceed to address
the issue of the chromomagnetic instability at equilibrium. Initially we shall
consider the case  $\Delta<<\delta$,
-the gap parameter much smaller than the displacement. This is
the situation near a second order phase transition
to the normal phase. 
For a LOFF state consisting of a single plane wave at $T=0$ this occurs 
at \cite{takada}
\begin{equation}
\delta =\delta_{c}\simeq 0.754\Delta_0,
\end{equation} 
which is also the maximum for the LOFF condensate, eq. (\ref{order}).
Although it has been argued that the single plane 
wave ansatz is not energetically
favorable compared with the multi plane wave ansatz in this neighbourhood 
and that the transition to the normal phase may be a first order one 
\cite{bowers}, we
shall still pursue the small 
gap expansion since it provides a full analytic treatment and
consequently can be very instructive.
We shall focus on the LOFF state at $T=0$.

Let's expand the free energy, eq. (\ref{freeenergy}) to 
quartic power of the small parameter, $\hat\Delta\equiv\Delta/\delta$,
\begin{equation}
\Gamma = \frac{2\bar\mu^2\delta^2}{\pi^2}\Big[\alpha\hat\Delta^2
+\frac{1}{2}\beta\hat\Delta^4\Big]
\end{equation}
with,
\begin{equation}
\alpha = \ln\frac{2\delta}{\Delta_0}-1+\frac{\rho+1}{2\rho}\ln(\rho+1)
+\frac{\rho-1}{2\rho}\ln|\rho-1|
\end{equation}
and
\begin{equation}
\beta = \frac{1}{4(\rho^2-1)}.
\end{equation}
The corresponding expansions of the quantities $A$ and $B$ read
\begin{equation}
A = \frac{g^2\bar\mu^2}{6\pi^2}(a_0\hat\Delta^2+a_1\hat\Delta^4+...),
\label{trans_47}
\end{equation}
\begin{equation}
B = \frac{g^2\bar\mu}{6\pi^2}(b_0\hat\Delta^2+b_1\hat\Delta^4+...),
\label{long_47}
\end{equation}
where
\begin{equation}
a_0 = \frac{3}{8\rho^3}\Big(2\rho-\ln\frac{\rho+1}{\rho-1}\Big),
\end{equation}
\begin{equation}
a_1 = -\frac{1}{16\rho^3}\Big(\frac{1}{\rho-1}+\frac{1}{\rho+1}\Big)
-\frac{1}{32\rho^3}\Big[\frac{1}{(\rho-1)^2}-\frac{1}{(\rho+1)^2}\Big],
\end{equation}
\begin{equation}
b_0 = \frac{3}{4\rho^2}\Big(\frac{2-\rho^2}{\rho^2-1}
+\frac{1}{\rho}\ln\frac{\rho+1}{\rho-1}\Big)
\end{equation}
and
\begin{equation}
b_1 = -\frac{1}{8\rho^3}\Big(\frac{1}{\rho-1}+\frac{1}{\rho+1}\Big)
-\frac{1}{16\rho^3}\Big[\frac{1}{(\rho-1)^2}+\frac{1}{(\rho+1)^2}\Big]
+\frac{1}{16\rho}\Big[\frac{1}{(\rho+1)^3}+\frac{1}{(\rho-1)^3}\Big]
\end{equation}
Similarly, we find
\begin{equation}
C = \frac{g^2\bar\mu^2}{9\pi^2}\Big(1+\frac{e^2}{3g^2}\Big)
(c_0\hat\Delta^2+c_1\hat\Delta^4+...),
\label{trans_8}
\end{equation}
\begin{equation}
D = \frac{g^2\bar\mu^2}{9\pi^2}\Big(1+\frac{e^2}{3g^2}\Big)
(d_0\hat\Delta^2+d_1\hat\Delta^4+...).
\label{long_8}
\end{equation}
where 
\begin{equation}
c_0 = \frac{3}{4\rho^3}\Big(2\rho-\ln\frac{\rho+1}{\rho-1}\Big) = 2a_0,
\end{equation}
\begin{equation}
c_1 = \frac{3}{32\rho^3}\Big[-\frac{2}{\rho+1}
-\frac{2}{\rho-1}+\frac{1}{(\rho+1)^2}-\frac{1}{(\rho-1)^2}\Big],
\end{equation}
\begin{equation}
d_0 = \frac{3}{2\rho^2}\Big(\frac{2-\rho^2}{\rho^2-1}
+\frac{1}{\rho}\ln\frac{\rho+1}{\rho-1}\Big)=2b_0
\end{equation}
and
\begin{equation}
d_1 = \frac{3}{16\rho^3}\Big[\frac{3}{\rho+1}+\frac{3}{\rho-1}
-\frac{3}{(\rho+1)^2}+\frac{3}{(\rho-1)^2}
+\frac{1}{(\rho+1)^3}+\frac{1}{(\rho-1)^3}\Big].
\end{equation}

The validity of the relations $c_0=2a_0$ and $d_0=2b_0$ is not  
restricted to zero temperature and the weak coupling approximation 
employed here. 
If we expand the integrand of the Meissner mass tensor (\ref{eqkit1})
to the order $\hat\Delta^2$, we find that the loop integrals for $a_0$ and 
$b_0$ are related to those for $c_0$ and $d_0$ through integration 
by parts \cite{giannak}. The same argument also leads to identical criteria for
the chromomagnetic instability of the 4-7th gluons and the 8th gluon 
near $T_c$ for g2SC. Minimization of the free energy with respect to $\Delta$ 
and $\rho$ yields
\begin{equation}
\hat\Delta^2 = 4(\rho_c^2-1)\Big(1-\frac{\delta}{\delta_c}\Big),
\end{equation}
and
\begin{equation}
\rho = \rho_c+\frac{1}{4}\frac{\rho_c}{\rho_c^2-1}\hat\Delta^2,
\end{equation}
where $\rho_c\simeq 1.20$ is the solution of the equation \cite{takada}
\begin{equation}
\frac{1}{2\rho_c}\ln\frac{\rho_c+1}{\rho_c-1}=1
\end{equation}
and
\begin{equation}
\frac{\delta_c}{\Delta_0}=\frac{1}{2\sqrt{\rho_c^2-1}}\simeq 0.754.
\end{equation}
The free energy at the minimum reads
\begin{equation}
\Gamma_{min}=-\frac{4{\bar\mu}^2}{\pi^2}(\rho_{c}^2-1)(\delta-\delta_c)^2
\end{equation}
and both branches of eq. (\ref{spectrum}) are gapless here.
The solution
satisfies the stability
condition (\ref{stability}). Substituting it into
eqs.(\ref{trans_47}), (\ref{long_47}),
(\ref{trans_8}) and (\ref{long_8}), we obtain
\begin{equation}
A=\frac{g^2\bar\mu^2}{96\pi^2}\frac{\hat\Delta^4}
{(\rho_c^2-1)^2} \ge 0,
\qquad B=\frac{g^2\bar\mu^2}{8\pi^2}\frac{\hat\Delta^2}{\rho_c^2-1}\ge 0
\end{equation}
and
\begin{equation}
C = 0, \qquad D=\frac{g^2\bar\mu^2}{6\pi^2}\Big(1+\frac{e^2}{3g^2}\Big)
\frac{\hat\Delta^2}{\rho_c^2-1}\ge 0
\end{equation}
Therefore the chromomagnetic instabilities are
removed completely for small gap parameters.

The situation for a multi-plane wave ansatz for
the LOFF state will be better if the small 
gap expansion remains numerically approximate. As we have seen for the single 
plane wave ansatz, the 
Meissner tensors are purely longitudinal to 
$\Delta^2$ and the logitudinal components are positive at the equilibrium. It 
follows that if the LOFF state (\ref{order_general}) contains 
three linearly independent momenta,
the Meissner tensor will
be positive definite, since it is additive with respect to different
terms of eq. (\ref{order_general}) to order $\Delta^2$.

\section{Numerical Results}

In the last section, we have shown analytically that the chromomagnetic 
instability is absent in the region
near the outer edge of the LOFF window at 
zero temperature. In this section, we shall explore 
the interior region in order to determine if the absence of 
the chromomagnetic instability 
can be extended to the entire LOFF region at $T=0$.

Before presenting our numerical results, we shall emphasize some general 
features of the solutions of eqs. (\ref{equil_1}) and (\ref{equil_2}). 
A trivial solution to eq. (\ref{equil_2}), $\rho=0$, corresponds to a 
homogeneous 2SC and there is always a solution to eq.(\ref{equil_1})
with $\Delta=\Delta_0$ for $0<\delta<\Delta_0$. The expression for
the free energy then becomes
\begin{equation}
\Gamma = \frac{2\bar\mu^2}{\pi^2}\Big(\delta^2-\frac{1}{2}\Delta_0^2\Big).
\label{free_bcs}
\end{equation}
Since the excitation spectrum, eq.
(\ref{spectrum}), is fully gapped in this case, we shall 
refer to the 
correponding state as the BCS state even in the presence of a
displacement, 
following the convention of Ref. \cite{takada}. We observe that the 
BCS state becomes meta-stable relative to the normal phase for 
\begin{equation}
\delta>\frac{1}{\sqrt{2}}\Delta_0.
\label{threshold}
\end{equation}
For the displacement within the interval $\frac{\Delta_0}{2}<\delta<\Delta_0$, 
there is another solution at $\rho=0$ with
\begin{equation}
\Delta=\sqrt{\Delta_0(2\delta-\Delta_0)}
\end{equation} 
which leads to the following expression for the free energy
\begin{equation}
\Gamma=\frac{\bar\mu^2}{\pi^2}(2\delta-\Delta_0)^2>0.
\end{equation}
This solution corresponds to the Sarma state that
is unstable in the grand canonical ensemble for a given $\delta$. 
In a canonical ensemble, it may be the only charge neutral state for certain 
pairing strength \cite{shovkovy} but suffers from chromomagnetic 
instabilities \cite{huang}. 

In what follows we shall restrict the terminology 'LOFF state' to
the nontrivial solution to eq. (\ref{equil_2}) 
that satisfies
\begin{eqnarray}
1 & - & \frac{(\rho+1)^2}{8\rho^3}\Big[3(1-x_1^2)\ln\frac{1+x_1}{1-x_1}
+4(\rho+1)x_1^3-6x_1\Big]\nonumber \\
& - & \frac{(\rho-1)^2}{8\rho^3}\Big[-3(1-x_2^2)\ln\frac{1+x_2}{1-x_2}
+ 4(\rho-1)x_2^3+6x_2\Big]=0
\label{equil_3}
\end{eqnarray}
together with eq. (\ref{equil_1}). It follows
from eq.(\ref{equil_1}) that
\begin{equation}
 \Delta<\Delta_0
\label{gap_sol}
\end{equation}
at the solution and eq.(\ref{equil_3}) implies that   
\begin{equation}
\frac{\Delta}{\delta}<1+\rho.
\label{loff_region}
\end{equation}
Otherwise, $x_1=x_2=0$ and eq.(\ref{equil_3}) cannot hold.
Physically, eq.(\ref{gap_sol}) is the consequence of the reduced phase 
space available for pairing and eq.(\ref{loff_region}) is the condition 
for the existence of gapless excitations, which is
necessary to make the net baryon number 
current vanish. Combining eq. (\ref{gap_sol}) and (\ref{loff_region}), we find 
$\delta<\Delta_0$ for the LOFF state.
 
For a small gap parameter, an analytical solution was obtained 
in the last section, which corresponds to a stable state
with lower energy compared with the BCS and the normal
states and is free of chromomagnetic instabilities. For a general 
gap parameter, however,
eqs. (\ref{equil_1}) and (\ref{equil_3}) can only be solved numerically. 
Our strategy is to find solutions of (\ref{equil_3})
for $\Delta/\delta$ for a given $\rho$ and
subsequently  to calculate the corresponding 
$\delta/\Delta_0$ in terms of (\ref{equil_1}). Concurrently, we examine the 
local stability according to eqs. (\ref{stability}) and the competition 
between
the LOFF state we find and the BCS state of the same $\delta/\Delta_0$. The 
Meissner tensors are calculated as well. 
We observe that the quantities $\rho$, ${\delta}/{\Delta_0}$, the
free energy difference, the Meissner masses and the stability determinant
of eq. {\ref{stability}), at the LOFF solution are single-valued
functions of the normalized gap parameter ${\Delta}/{\Delta_0}$. These
functions are plotted in Figs. 1-7.

The diquark momentum of a LOFF state, normalized by 
the displacement parameter versus 
the normalized gap parameter is shown in Fig.1 and the 
corresponding ratio of the 
displacement parameter to the BCS gap is shown in Fig.2. In Fig.3, we 
display the free energy difference between a LOFF state and a BCS state 
at the same $\delta$, normalized by the magnitude of the BCS energy (\ref{free_bcs})
at $\delta=0$.
Combining the three figures, we find the LOFF window, in which the LOFF 
state is of the lowest energy, 
\begin{equation}
\delta_{\rm min}<\delta<\delta_{\rm max},
\label{window}
\end{equation} 
where $\delta_{\rm max}=\delta_c\simeq 0.754\Delta_0$ and 
$\delta_{\rm min}\simeq 0.706\Delta_0$, slightly 
below the threshold (\ref{threshold}). The corresponding momentum
of the diquark increases monotonically from $\rho\simeq 1.20$ 
at $\delta_{\rm max}$ to
$\rho\simeq 1.28$ at $\delta_{\rm min}$.
Our results confirm those of Ref. \cite{takada}.
In addition, we found that the normalized gap parameter,
$\Delta/\Delta_0$ varies monotonically from 0 at $\delta_{\rm max}$
to 0.242 at $\delta_{\rm min}$. So the LOFF window on all
figures corresponds to the domain $0<\Delta/\Delta_0<0.242$
of the abscissa.

The numerical results of the Meissner tensors, Figs. 4-6 are most interesting. 
The numbers shown are normalized by the Meissner mass square of the BCS state 
at $\delta=0$, i.e.
\begin{equation}
(m^2)_{ij}=A_0\delta_{ij}
\end{equation}
and 
\begin{equation}
(m^{\prime 2})_{ij}=D_0\delta_{ij}
\end{equation}
with 
\begin{equation}
A_0=\frac{g^2\bar\mu^2}{6\pi^2}, \quad D_0=\frac{g^2\bar\mu^2}{9\pi^2}
\Big(1+\frac{e^2}{3g^2}\Big).
\end{equation}  
All squared Meissner tensors are nonnegative within the LOFF window and 
the region free from the chromomagnetic instabilities extends beyond it. While 
the Meissner tensor associated to the eighth gluon is aways logitudinal, 
Fig. 6, the transverse
component of the Meissner tensor associated to the 4-7th gluons, Fig.4 
is aways much 
smaller than its logitudinal component, Fig.5. The stability of the solution 
against virtual displacements of $\Delta$ and $\rho$ within the LOFF window is 
guaranteeted by the positivity of $D$, plotted in Fig. 6 and the positivity of 
the determinant in eq.(\ref{stability}), plotted in Fig. 7.
The non-smooth behavior of the function $D$ and the determinant of Figs.
6 and 7 correspond to the transition of the LOFF state from the region
where both branches of (\ref{spectrum}) are gapless, $x_2>0$, 
to the region where only the $E_{\vec p}^{-}$ branch is gapless, $x_2=0$. 
The derivative of $x_2$ with respect to both $\Delta$ and
$\rho$ become singular there. When the gap parameter is expressed in
terms of $\rho$ and $x_2 > 0$ according to eq. (\ref{chi2}) and the
function $D$ and the determinant are expanded according to the powers
of $x_2$, the linear term is sensitive to the singularity and gives
rise to the non-smooth behavior. On the other hand, the same type
of expansion of the functions $A$ and $B$ does not contain a linear
term in $x_2$ and thus the transition behavior of $A$ and $B$ is much
more smooth.

As the LOFF momentum goes to zero, $\rho\to 0$, we have $\Delta\to\Delta_0$ 
and $\delta\to\Delta_0$, a result can be established analytically. Therefore 
the LOFF state, the BCS state and the Sarma state joins at the point 
( $\Delta_0$, $\Delta_0$ 0 ) in the three dimensional parameter space of 
$\Delta$, $\delta$ and $\rho$. We also observe from Figs. 4-7 that the
chromomagnetic instability and the instability against a small variation
of parameters show up for sufficiently low values of $\rho$. Therefore
the instability free region of a LOFF state is disconnected from the
g2SC state in the parameter space.

\section{Concluding Remarks}

In this paper, we have systematically examined the Meissner effect in a
two-flavor color superconductor with a LOFF pairing. We found 
that within the 
LOFF window at $T=0$, that is within the range of the displacement parameter 
where the LOFF state is energetically favored, all the squares of
the Meissner masses are 
non-negative. Although the positivity of the longitudinal component of the 
Meissner mass tensor associated with the eighth gluon 
follows from the stability 
of the LOFF state against a variation of the LOFF momentum, $\rho$, the 
positivity of the Meissner tensor for the 4-7th gluons appears
accidental regarding the pairing ansatz (\ref{order}). We have also found that 
the Meissner tensor of the 8th gluon is always longitudinal. This
is also the case with the Meissner effect in an electronic
LOFF superconductor.

The charge neutrality condition has to be implemented before
we can claim that the LOFF state is the ground state of quark matter
at moderately high baryon density. In the
previous letter \cite{giannak}, we showed that the
chromomagnetic instability of the 8th gluon pertaining to g2SC implies a
state of lower free energy with a nonzero diquark momentum and without
offsetting the charge neutrality. This is not sufficient since the LOFF
window, with the diquark momentum comparable to the Fermi momentum
displacement, is quite far away from the g2SC and is very narrow in the
three dimensional parameter space. It remains to be seen if quark matter
within the LOFF window can be made neutral. 
In a hypothetical world where the sum of the electric charges of the u
and d quarks is a small number, we can find within the
framework of the weak coupling approximation 
an analytic expression for an electrically neutral LOFF
phase that is located near the upper edge of 
the LOFF window and is energetically
favored over both the neutral normal phase and the neutral BCS phase.
Also it is free
of chromomagnetic instabilities. The details will be presented in
a forthcoming publication \cite{future}.
Notice
that the neutral phases being compared do not have the same displacement
parameter $\delta$ since the charge neutrality constraint adjusts
$\delta$ differently for different phases. In the real world, the sum of
the charges of the u and d quarks is $1/3$. The weak coupling
approximation may be marginal and one may have to seek numerical
solutions of eq. (\ref{equil_1}), eq.(\ref{threshold}) 
and the charge neutrality constraint
simultaneously.

We also notice that, if the pairing force is mediated by
the long range one-gluon exchange of QCD
instead of the point coupling of NJL model, the
LOFF window will be at least five times wider under the same $\Delta_0$
($\delta_{\rm max}\simeq 0.968\Delta_0$ and
$\delta_{\rm min}\le 0.706\Delta_0$)\cite{GLR}, making
the implementation of the charge neutrality more likely.

In this paper we have considered only the simplest LOFF
ansatz that consists of one plane wave and pairs $u$ and $d$ quarks. It
was suggested that multi plane wave condensates are favored more near the
upper edge of the LOFF window \cite{bowers}. Numerical
analysis also suggests that the $s$ quark cannot be ignored in pairing at
$T=0$. A more complicated structure of the condensate is required for a
realistic LOFF pairing in quark matter and may not 
be accessible via analytical means. In
any case, the absence of chromomagnetic instabilities for the simple
2SC-LOFF is encouraging and promotes the LOFF state as a potential
candidate of the ground state of quark matter at moderately baryon
densities.

\section*{Acknowledgments}
Hai-cang Ren thanks DeFu Hou for valuable discussions.
We are indebted to I. Shovkovy for an interesting communication.
This work is supported in 
part by the US Department of Energy under grants
DE-FG02-91ER40651-TASKB.

\section*{Appendix A}

In this appendix, we shall first derive the exact expression of the quark 
propagator in the superconducting phase for $\delta^\prime=0$. 
Subsequently we shall perform the
weak coupling expansion. Finally, we shall 
approximate the determinant of the quark propagator which enters 
the free energy function (\ref{gamma}) in a similar manner.

As $\tau_3$ and $\lambda_2$ are the only isospin and
color matrices appearing in the inverse quark propagator (\ref{prop_gap}) 
at $\delta^\prime=0$, they can be treated as a c-numbers. Taking the 
square of the inverse propagator, we have
\begin{eqnarray}
[{\cal S}^{-1}(P)]^2 &=& (i\nu-\delta\tau_3)^2+\bar\mu^2-p^2-q^2
-\Delta^2\lambda_2^2\\ \nonumber
&+&2\rho_3[\bar\mu(i\nu-\delta\tau_3)-\vec p\cdot\vec q]
+2\Delta\rho_1\gamma_5(-i\vec\gamma\cdot\vec q+\bar\mu\gamma_4)\lambda_2.
\end{eqnarray}
In terms of the matrix
\begin{eqnarray}
N(P) &=& (i\nu-\delta\tau_3)^2+\bar\mu^2-p^2-q^2-\Delta^2\lambda_2^2
\nonumber \\
&-&2\rho_3[\bar\mu(i\nu-\delta\tau_3)-\vec p\cdot\vec q]
-2\Delta\rho_1\gamma_5(-i\vec\gamma\cdot\vec q+\bar\mu\gamma_4)\lambda_2,
\end{eqnarray}
we obtain the exact expression of the quark propagator
\begin{equation}
{\cal S}(P)=\frac{{\cal S}^{-1}(P)N(P)}{D(P)},
\label{exact}
\end{equation}
where
\begin{equation}
D(P)=[(i\nu-\delta\tau_3)^2+\bar\mu^2-p^2-q^2-\Delta^2\lambda_2^2]^2
-4[\bar\mu(i\nu-\delta\tau_3)-\vec p\cdot\vec q]^2
+4\Delta^2(\bar\mu^2-q^2)\lambda_2^2,
\end{equation}
and $\lambda_2^2$ projects into the subspace 
of the red and green colors.

In order to perform the weak coupling 
approximation, we regard $p-\bar\mu$, $\nu$,
$\delta$ and $q$ as small quantities of comparable magnitudes
and maintain only the leading order terms in the numerator and 
the denominator of (\ref{exact}). We find that
\begin{equation}
D(P)\simeq 4\bar\mu^2[-(i\nu-\delta\tau_3-\hat p\cdot\vec q)^2
+(p-\bar\mu)^2+\Delta^2\lambda_2^2]
\label{approx_D}
\end{equation}
and
\begin{equation}
{\cal S}^{-1}(P)N(P)\simeq -2\bar\mu^2(-i\vec\gamma\cdot\hat p+\rho_3\gamma_4)
[|\vec p-\rho_3\vec q|-\bar\mu+\rho_3(i\nu-\delta\tau_3)
+\Delta\lambda_2\gamma_5\gamma_4]
\label{approx_N}
\end{equation}
Upon substitution of (\ref{approx_D}) and (\ref{approx_N}) into 
(\ref{exact}), we derive the approximate quark propagator expression
(\ref{approx_S}) that appeared in section II.

The determinant of ${\cal S}^{-1}(P)$ may be calculated similarly.
Let's write
\begin{equation}
{\rm det}{\cal S}^{-1}(P) ={\rm det}^{\frac{1}{2}}[[{\cal S}^{-1}(P)]^2]
\simeq (2\bar\mu)^{24}{\rm det}^{\frac{1}{2}}d(P),
\end{equation}
where
\begin{equation}
d(P)=\bar\mu-p+\rho_3(i\nu-\delta\tau_3-\hat p\cdot\vec q)
+\Delta\lambda_2\rho_1\gamma_5\gamma_4.
\end{equation}
Now by decomposing the 48$\times$48 matrix $d(P)$, according to the 
eigenvalues of $\tau_3$ and $\lambda_2$, i.e.
\begin{equation}
{\rm det}d(P)=\prod{\rm det}d(P)\mid_{\tau_3=\pm 1,\lambda_2=\pm 1,0}
\end{equation}
we have
\begin{equation}
{\rm det}d(P)\mid_{\tau_3=\pm 1,\lambda_2=1}
=[(p-\bar\mu)^2+\Delta^2-(i\nu\mp\delta-\hat p\cdot\vec q)^2]^4,
\end{equation}
\begin{equation}
{\rm det}d(P)\mid_{\tau_3=\pm 1,\lambda_2=-1}
=[(p-\bar\mu)^2+\Delta^2-(i\nu\mp\delta-\hat p\cdot\vec q)^2]^4
\end{equation}
and
\begin{equation}
{\rm det}d(P)\mid_{\tau_3=\pm 1,\lambda_2=0}
=[(p-\bar\mu)^2-(i\nu\mp\delta-\hat p\cdot\vec q)^2]^4.
\end{equation}
Eq. (\ref{deter_approx}) is then established.

\section*{Appendix B}

The integrand of the
angular integrals for the free energy function (\ref{egw})
and that for the Meissner tensor, eqs. (\ref{HS4_7}) and (\ref{HS8}),  
contain terms with $\theta(|\delta_\theta|-\Delta)$. It
confines
the integral within the region in the phase space where 
either or both branches of the excitation spectrum, $E_{\vec p}^{\pm}$ 
of (\ref{spectrum}), become gapless. 
This occurs when $\Delta<(1+\rho)\delta$. Otherwise the integration 
domain specified by $\theta(|\delta_\theta|-\Delta)$ is null.
There are three different cases to be considered:

\noindent
1) $|1-\rho|\delta<\Delta<(1+\rho)\delta$:

Only the $E_{\vec p}^-$ branch may be gapless and we have 

\begin{equation}
\int_{-1}^1d\cos\theta\theta(|\delta_\theta|-\Delta)(...)
=\int_{-1}^{\frac{\delta-\Delta}{\rho\delta}}
d\cos\theta(...).
\end{equation}

\noindent
2) $\rho<1$ and $\Delta<(1-\rho)\delta$:

Again we may have only one gapless branch, $E_{\vec p}^-$, but instead
\begin{equation}
\int_{-1}^1d\cos\theta\theta(|\delta_\theta|-\Delta)(...)
=\int_{-1}^1
d\cos\theta(...).
\end{equation}
because $\frac{\delta-\Delta}{{\rho}{\delta}}>1$.

\noindent
3) $\rho>1$ and $(\rho-1)\delta<\Delta<(\rho+1)\delta$:

Both branches, $E_{\vec p}^\pm$ may be gapless and we have
 
\begin{equation}
\int_{-1}^1d\cos\theta\theta(|\delta_\theta|-\Delta)(...)
=\int_{-1}^{\frac{\delta-\Delta}{\rho\delta}}
d\cos\theta(...)+\int_{\frac{\delta+\Delta}{\rho\delta}}^1
d\cos\theta(...).
\end{equation}

The indefinite integrals of the integrands of the free energy 
function and the Meissner tensors are all elementary functions.

\section*{Appendix C}

In this appendix, we shall supply the details supporting the statement 
in section III that the identity (\ref{TPgamma5}) for the 
quark propagator follows from the invariance of the 2SC-LOFF 
under a combined 
transformation of a space inversion, a time reversal and a $\gamma_5$
multiplication. We shall work with the canonical formulation at 
$T=0$ though the generalization to $T\neq 0$ is straightforward.

The Hamiltonian operator corresponding to the mean field NJL action 
(\ref{NJL_MF}) reads
\begin{eqnarray}
H_{MF} &=& \Omega\frac{\Delta^2}{4G_D}+\int d^3{\vec r}[{
\bar\psi\vec\gamma\cdot\vec\nabla\psi-\bar\psi\gamma_4\mu\psi} \nonumber \\
& - & \Delta (-\bar\psi_C\gamma_5\lambda_2\tau_2\psi{e^{-2i{\vec q}{\vec r}}}
+\bar\psi\gamma_5\lambda_2\tau_2\psi_C{e^{2i{\vec q}{\vec r}}})].
\label{NJL_NP}
\end{eqnarray}
Introduce an anti-unitary operator ${\cal U}$ which implements a 
space inversion, a time reversal and a $\gamma_5$ multiplication. 
The transformation laws of the field operators in Heisenberg representation 
are given by
\begin{eqnarray}
{\cal U}\psi(\vec r,t){\cal U}^{-1} &=& i\gamma_2\psi(-\vec r,-t)\nonumber \\
{\cal U}\psi^\dagger(\vec r,t){\cal U}^{-1} &=& 
-i\psi^\dagger(-\vec r,-t)\gamma_2.
\end{eqnarray}
The corresponding transformations of the NG spinors are 
\begin{eqnarray}
{\cal U}\Psi(\vec r,t){\cal U}^{-1} &=& i\rho_3\gamma_2
\Psi(-\vec r,-t)\nonumber \\
{\cal U}\Psi^\dagger(\vec r,t)
{\cal U}^{-1} &=& -i\Psi^\dagger(-\vec r,-t)\gamma_2\rho_3,
\label{NG_law}
\end{eqnarray}
following (\ref{loff}) and (\ref{NGbasis}).
The phase factor associated to the time reversal has been 
fixed in accordance with 
phase convention of the condensate such that
\begin{equation}
{\cal U}H_{MF}{\cal U}^{-1}=H_{MF}.
\end{equation}
For a state $|>$ that is invariant under ${\cal U}$, say the 
ground state and an arbitrary 
operator $O$, we have \cite{tdl}
\begin{equation}
<|O^\dagger|>=<|{\cal U}O{\cal U}^{-1}|>.
\label{identity}
\end{equation}
The Hamiltonian (\ref{NJL_NP}) is also invariant under a unitary 
transformation, $U$, that generates a translation and a rephasing 
according to 
\begin{eqnarray}
U\psi(\vec r,t)U^\dagger &=& e^{-i\vec q\cdot\vec a}
\psi(\vec r+\vec a,t)\nonumber \\
U\psi^\dagger(\vec r,t)U^\dagger &=& e^{i\vec q\cdot\vec a}
\psi^\dagger(\vec r+\vec a,t).
\label{trans}
\end{eqnarray}

Consider the quark propagator in coordinate space,
\begin{equation} 
{\cal S}_{\alpha\beta}(\vec r,t)
=<|\Psi_\alpha(\vec r,t){\bar\Psi}_\beta(0,0)|>
\theta(t)-<|{\bar\Psi}_\beta(0,0)\Psi_\alpha(\vec r,t)|>\theta(-t).
\end{equation}
It follows from (\ref{NG_law}), (\ref{identity}), (\ref{trans}) and time 
translation invariance that
\begin{equation}
{\cal S}_{\alpha\beta}(\vec r,t)=-({\rho_3C})_{\alpha\gamma}
{\cal S}_{\sigma\gamma}(\vec r,t)({C\rho_3})_{\beta\sigma},
\end{equation}
which, upon a Fourier transformation, gives rise to the identity 
(\ref{TPgamma5}).

\begin{figure}[t]
\epsfxsize 10cm
\centerline{\epsffile{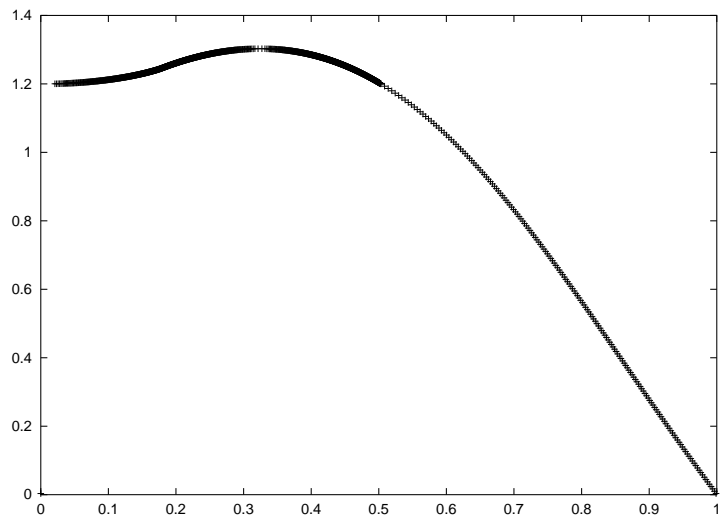}}
\medskip
\caption{The dimensionless diquark momentum $\rho$ of the LOFF state as
a function of the normalized gap parameter ${\Delta}/{\Delta_0}$.}
\end{figure}

\begin{figure}[t]
\epsfxsize 10cm
\centerline{\epsffile{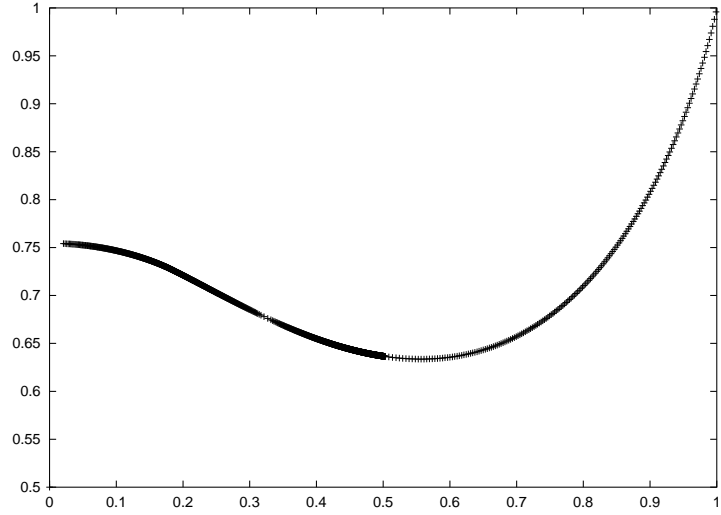}}
\medskip
\caption{The normalized displacement parameter, $\delta/\Delta_0$, of
a LOFF state as a function of the normalized gap parameter
$\Delta/\Delta_0$.}
\end{figure}

\begin{figure}[t]
\epsfxsize 10cm
\centerline{\epsffile{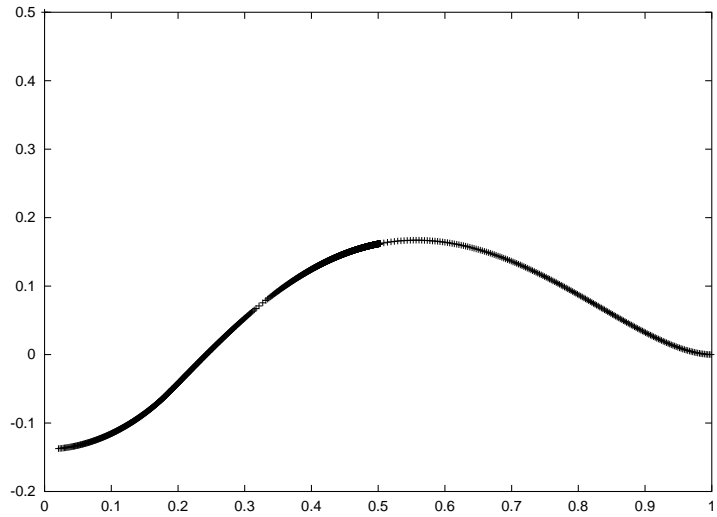}}
\medskip
\caption{The free energy difference between a LOFF state and a
BCS state as a function of the normalized gap parameter
$\Delta/\Delta_0$. The coordinate is normalized to the magnitude of the
free energy of the BCS state, eq.(94), at $\delta=0$ and the LOFF state
is favored when the value is negative.}
\end{figure}

\begin{figure}[t]
\epsfxsize 10cm
\centerline{\epsffile{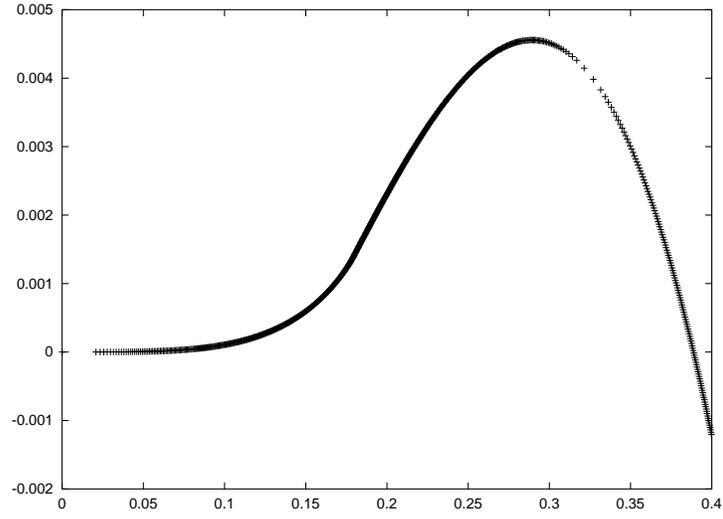}}
\medskip
\caption{The normalized transverse Meissner mass square of the
4-7 gluons, $A/A_0$, as a function of the normalized gap parameter
$\Delta/\Delta_0$ of a LOFF state.}
\end{figure}

\begin{figure}[t]
\epsfxsize 10cm
\centerline{\epsffile{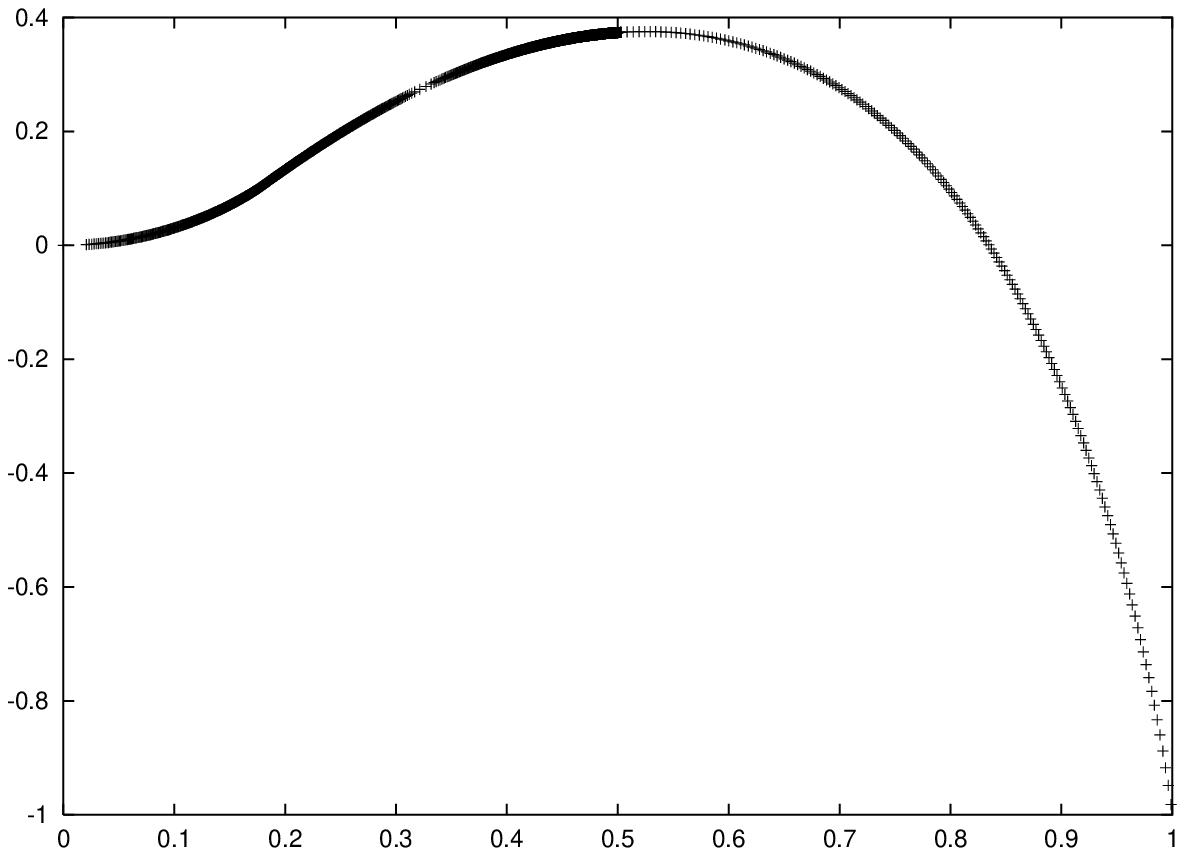}}
\medskip
\caption{The normalized logitudinal Meissner mass square of the
4-7th gluons, $B/A_0$, as a function of the normalized gap parameter
$\Delta/\Delta_0$ of a LOFF state.}
\end{figure}

\begin{figure}[t]
\epsfxsize 10cm
\centerline{\epsffile{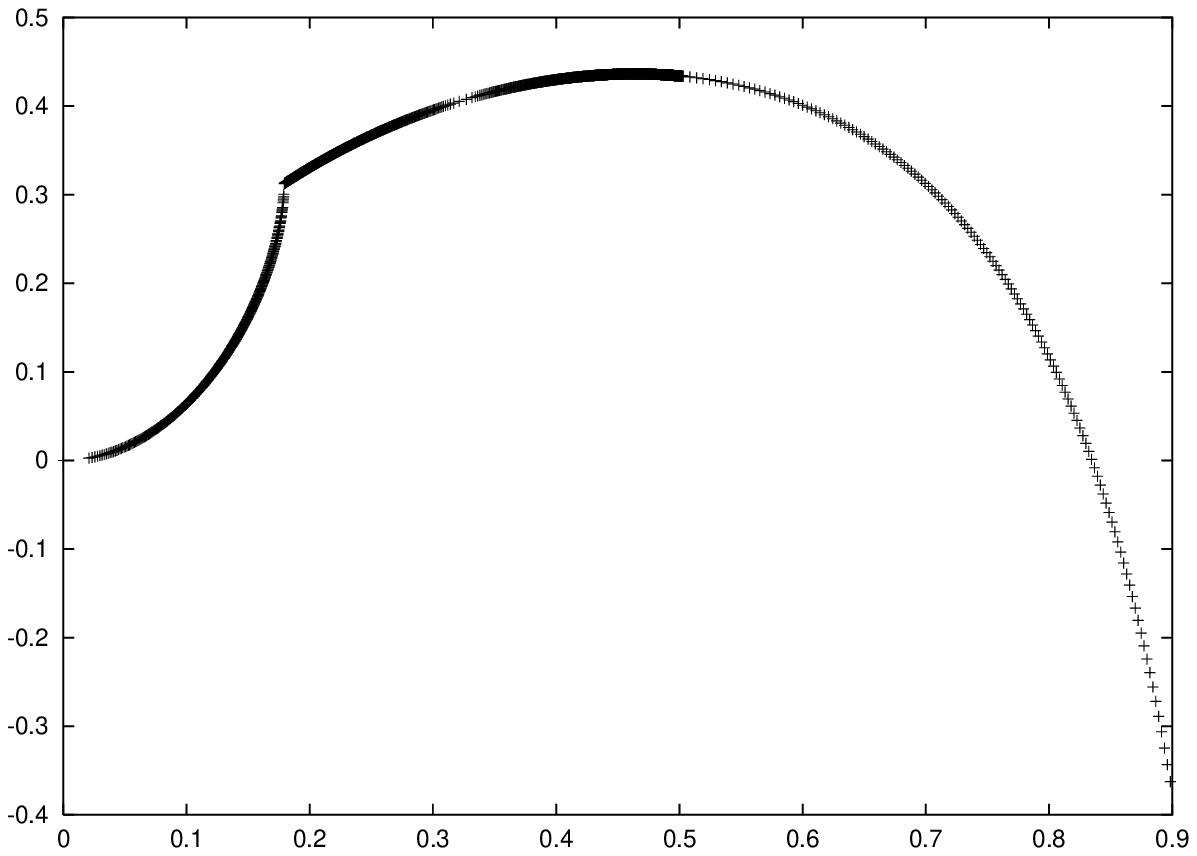}}
\medskip
\caption{The normalized logitudinal Meissner mass square of the
8th gluon, $D/D_0$, as a function of the normalized gap parameter
$\Delta/\Delta_0$ of a LOFF state.}
\end{figure}

\begin{figure}[t]
\epsfxsize 10cm
\centerline{\epsffile{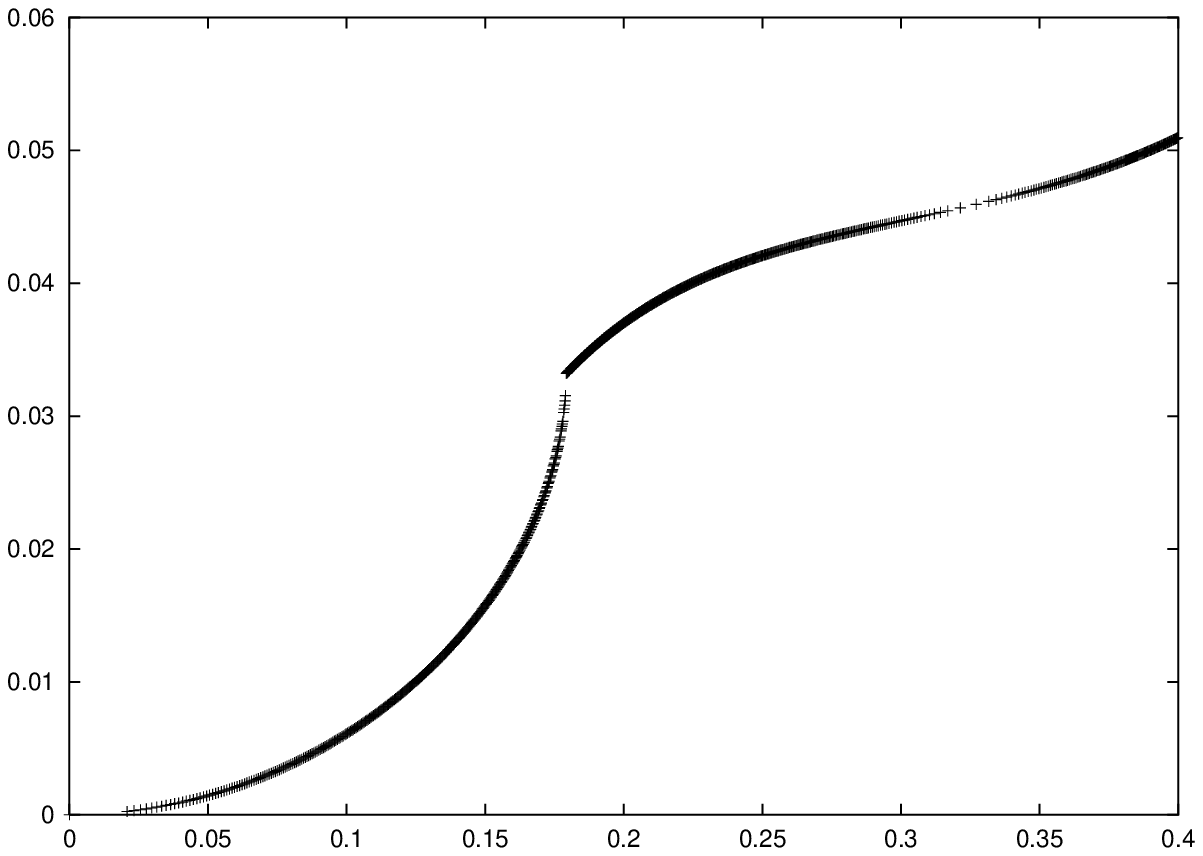}}
\medskip
\caption{The stability determinant in eq.(13)(arbitrary unit) of
a LOFF state as a function of the normalized gap parameter
$\Delta/\Delta_0$.}
\end{figure}


\begin{thebibliography}{99}
\bibitem{loff}
A. Larkin and Yu. N. Ovchinnikov, Zh. Eksp. Teor. Fiz. {\bf 47}, 1136 
(1964) ( Soviet Physics, JETP, {\bf 20}, 762 (1965); P. Fulde and R. A. 
Ferrell, Phys. Rev. {\bf 135}, A550 (1964).
\bibitem{takada} S. \ Takada and T. \ Izuyama, Prog. \ Theor. \
Phys. {\bf 41} (1969) 635.
\bibitem{reviews} K.\ Rajagopal and F.\ Wilczek, 
in B.L.\ Ioffe Festschrift, {\it At the Frontier of Particle
Physics/Handbook of QCD}, M.\ Shifman ed., 2061 (World Scientific 2001);
M.\ Alford, Ann.\ Rev.\ Nucl.\ Part.\ Sci.\ {\bf 51} (2001) 131;
T.\ Sch\"afer, hep-ph/0304281; D. H. Rischke, Prog. Part. Nucl. Phys., 
{\bf 52}, 197 (2004);
H.-c.\ Ren, hep-ph/0404074, M. \ Huang, hep-ph/0409167;
I. \ Shovkovy, nucl-th/0410091 and references therein.
\bibitem{rapp} R. Rapp, T. Schaefer, E. V. Shuryak and M. Velkovsky, Phys. 
Rev. Lett., {\bf 81}, 53 (1998).
\bibitem{rajagopal} M.\ G.\ Alford, J.\ A.\ Bowers 
and K.\ Rajagopal, Phys.\ Rev.\ {\bf D63} (2001)
074016; A.\ K.\ Leibovich, K.\ Rajagopal
and E.\ Shuster, Phys. \ Rev. \ {\bf D64} (2001) 094005; 
R. \ Casalbuoni, G.\ Nardulli, Rev.\ Mod.\ Phys., {\bf 76}, (2004) 263.
\bibitem{GLR} I. \ Giannakis, J. \ T. \ Liu and H. \ C. \ Ren,
Phys.\ Rev.\ {\bf D66} (2002) 031501.
\bibitem{shovkovy} I.\ Shovkovy 
and M.\ Huang, Phys.\ Lett.\ {\bf B564} (2003) 205; 
M. Huang and I. Shovkovy, Nucl. Phys. {\bf A729} (2003), 835;
M. \ Huang, P. \ F. \ Zhuang 
and W. \ Q. \ Chao, Phys. \  Rev.  {\bf D67} (2003) 065015;
Jinfeng Liao and Pengfei
Zhuang, Phys.\ Rev. \ {\bf D68} (2003) 114016;
A. Mishra and H. Mishra, Phys. Rev. {\bf D69}
(2004), 014014.
\bibitem{alford}
M. \ Alford, C. \ Kouvaris and K. \ Rajagopal, Phys. \ Rev. \ Lett.
{\bf 92} (2004) 222001.
\bibitem{phase}
K. Iida, T. Matsuura, M. Tachibana and
T. Hatsuda, Phys. \ Rev. \ Lett. {\bf 93}
(2004) 132001;
S. \ Ruester, I. \ Shovkovy
and D. \ Rischke, Nucl. \ Phys. {\bf A743}, (2003) 127;
S. Ruester, V. Werth, M. Buballa, I. Shovkovy and D. Rischke, 
{\it ''The phase diagram of neutral quark matter: self-consistent treatment 
of quark masses''}, hep-ph/0503184; D. Blaschke, S. Freriksson, H. Grigorian,
A. M. Oztas and F. Sandin, {\it ''The phase diagram of three-flavor quark 
matter under compact star constraint'' }, hep-ph/0503194;
Pengfei Zhuang, {\it ''Phase structure of color superconductivity''},
hep-ph/0503250; H. \ Abuki, M. \ Kitazawa and T. \ Kunihiro,
{\it How does the dynamical chiral condensation affect the three flavor
neutral quark matter?}, hep-ph/0412382.
\bibitem{sarma} G.\ Sarma, \ Phys. \ Chem. \  Solid. {\bf 24} (1963) 1029.
\bibitem{wilczek} W. V. \ Liu 
and F.\ Wilczek, \ Phys. \ Rev. \ Lett. {\bf 90} (2003)
047002; E. Gubankova, W. V. Liu and
F. Wilczek, Phys. \ Rev. \ Lett. {\bf 91}
(2003) 032001.
\bibitem{bedaque} P.\ F.\ Bedaque,
H. \ Caldas and G. \ Rupak, Phys. \ Rev. \ Lett. {\bf 91} (2003)
247002; H. \ Caldas, Phys. \ Rev. {\bf A69} (2004) 063602;
S. \ Reddy and G. \ Rupak, Phys. \ Rev. \ {\bf C71}
(2005) 025201.
\bibitem{huang} M.\ Huang and I.\ Shovkovy, Phys. \ Rev. \ {\bf D70}
(2004) 094030;
M.\ Huang and I.\ Shovkovy, 
Phys.\ Rev.\ {\bf D70} (2004) 051501(R).
\bibitem{casalbuoni} R.\ Casalbuoni, R.\ Gatto, M.\ Mannarelli, G.\ Nardulli
and M.\ Raggieri, Phys. \ Lett. {\bf B605} (2005) 362.
\bibitem{giannak} I. Giannakis and H. C. Ren, Phys. Lett. {\bf B611} (2005)
137.
\bibitem{AlfordWang} M. Alford, Q. H. Wang, {\it '' Photons in gapless color-
flavor locked quark matter''}. hep-ph/0501078.
\bibitem{hzc}
M. \ Huang, P. \ F. \ Zhuang and W. \ Q. \ Chao, Phys. \ Rev. {\bf D65} (2002)
076012.
\bibitem{mixing} M.\ Alford, J.\ Berges and K.\ Rajagopal, Nucl.\ Phys. 
\ {\bf B571}(2000); E. V. Gorbar, Phys. Rev. {\bf D62}(2000), 014007.
\bibitem{book} E. M. Lifshitz and L. P. Pitaevskii, {\it Statistical Physics},
(Bergamon Press) 1980, Chapter V.
\bibitem{future} I. Giannakis and Hai-Cang Ren, work in progress.
\bibitem{bowers} J. A. Bowers and K. Rajagopal, Phys. Rev. {\bf D66} (2002), 
065002.
\bibitem{tdl} T. D. Lee, {\it Particle Physics and Introduction to Field 
Theory }, Harwood Academic Publishers, 1981, Chapter 13. 

\end{thebibliography}
\end{document}